\documentclass{PoS}

\title{Non-Abelian Higgs Theory in a Strong Magnetic Field
and Confinement}

\ShortTitle{Non-Abelian Higgs Theory in a Strong Magnetic Field
and Confinement}

\author{\speaker{Hideo Suganuma}, \\
        Department of Physics \& Division of Physics and Astronomy, 
Graduate School of Science, \\
Kyoto University, 
Kitashirakawaoiwake, Sakyo, Kyoto 606-8502, Japan\\
        E-mail: \email{suganuma@scphys.kyoto-u.ac.jp}}

\abstract{
The non-abelian Higgs (NAH) theory is studied in a strong magnetic field. 
For simplicity, we study the SU(2) NAH theory with the Higgs triplet 
in a constant strong magnetic field $\vec B$, 
where the lowest-Landau-level (LLL) approximation can be used. 
Without magnetic fields, 
charged vector fields $A_\mu^\pm$ have a large mass $M$ 
due to Higgs condensation, 
while the photon field $A_\mu$ remains to be massless.
In a strong constant magnetic field 
near and below the critical value $eB_c \equiv M^2$, 
the charged vector fields $A_\mu^\pm$ 
behave as 1+1-dimensional quasi-massless fields, 
and give a strong correlation
along the magnetic-field direction 
between off-diagonal charges coupled with $A_\mu^\pm$. 
This may lead a new type of confinement 
caused by charged vector fields $A_\mu^\pm$.
}

\FullConference{The XIIIth Quark Confinement and the Hadron Spectrum, \\
31 July - 6 August 2018\\
Maynooth University, Maynooth, Ireland}

\begin{document}

\section{Introduction}

In the Weinberg-Salam model, through the Brout-Englert-Higgs mechanism 
\cite{EBH64}, all the weak bosons ($W_\mu^\pm$ and $Z_\mu$) acquire a large mass 
and can propagate to very short distance, 
and only photon remains massless and can propagate to long distance. 

Also for quantum chromodynamics (QCD), in the context of the dual superconductor picture 
for quark confinement \cite{N74,tH81,EI82}, 
by taking the maximally Abelian (MA) gauge, 
off-diagonal gluons acquire a large effective mass of about 1GeV and 
become infrared inactive \cite{AS99}, 
and only diagonal gluons contribute to long-distance physics, 
which is called ``Abelian dominance'' \cite{tH81,EI82,AS99,SY90,SS1415}.

In fact, QCD in the MA gauge and the non-Abelian Higgs (NAH) theory 
such as the Weinberg-Salam model are similar from the viewpoint of 
large mass generation of off-diagonal gauge bosons $A_\mu^\pm$, 
and sequential off-diagonal inactiveness and low-energy Abelianization, 
although the NAH theory does not exhibit charge confinement.

However, this situation can be drastically changed
in the presence of a strong magnetic field for 
the NAH theory, as will be discussed.
In this paper, we study the NAH theory in a strong magnetic field 
and consider a new type of confinement 
caused by charged vector fields $A_\mu^\pm$.
We here note that 
``magnetic properties of quantum systems'' or
``quantum systems in external magnetic fields'' are  
also interesting subjects in various fields in physics 
\cite{S77,NO78,AO89,ST91,M96,KMW08,CDV13,BCGK14}.

\section{SU(2) Non-Abelian Higgs Theory with Higgs Triplet}

We start from the standard SU(2) non-Abelian Higgs (NAH) theory 
with the SU(2) gauge field $A_\mu^{^{\rm SU(2)}} =A_\mu^a T^a \in su(2)$ and 
the SU(2) Higgs-scalar triplet $\Phi^a$ ($a=1,2,3$),
\begin{eqnarray}
{\cal L}_{\rm NAH} = -\frac{1}{4}G_{\mu\nu}^a G^{\mu\nu}_a
+\frac{1}{2}D_{\mu}^{^{\rm SU(2)}} \Phi^a 
D^{\mu}_{_{\rm SU(2)}} \Phi^a-\frac{\lambda}{4}(\Phi^a\Phi^a-v^2)^2,
\end{eqnarray}
where the SU(2) covariant derivative $D_\mu^{^{\rm SU(2)}} \equiv \partial_\mu
+ieA_\mu^{^{\rm SU(2)}}$ satisfies  
$D_\mu^{^{\rm SU(2)}} \Phi^a =\partial_\mu \Phi^a-e\epsilon^{abc}A_\mu^b\Phi^c$, 
and the SU(2) field strength $G_{\mu\nu}
\equiv\frac{1}{ie}[A_\mu^{^{\rm SU(2)}},~A_\nu^{^{\rm SU(2)}}]$ 
is written as $G_{\mu\nu}^a=\partial_\mu A_\nu^a-\partial_\nu A_\mu^a
-e\epsilon^{abc}A_\mu^b A_\nu^c$.

At the tree level, the Higgs field $\Phi^a$ has a vacuum expectation value, 
and one can set $\langle \Phi^3 \rangle = v~(\ge 0) 
\in {\bf R}$ and $\Phi^3=v+\sigma$ in the unitary gauge. Then, one obtains  
\begin{eqnarray}
{\cal L}_{\rm NAH}=&-&\frac{1}{4}[F_{\mu\nu}-ie(A_\mu^+A_\nu^--A_\mu^-A_\nu^+)]^2
-\frac{1}{2}(D_\mu^-A_\nu^+-D_\nu^-A_\mu^+)(D^\mu_+ A^\nu_--D^\nu_+ A^\mu_-) \cr
&+&(M+e\sigma)^2 A_\mu^+A^\mu_- 
+\frac{1}{2}(\partial_\mu\sigma)^2-\frac{\lambda}{4} \{(v+\sigma)^2-v^2\}^2, \cr
=&-&\frac{1}{4}F_{\mu\nu} F^{\mu\nu}
-\frac{1}{2}(D_\mu^-A_\nu^+-D_\nu^-A_\mu^+)(D^\mu_+ A^\nu_--D^\nu_+ A^\mu_-)
+(M+e\sigma)^2 A_\mu^+A^\mu_- \cr
&+&ieF^{\mu\nu}A_\mu^+A_\nu^- 
+\frac{1}{2}e^2[(A_\mu^+A^\mu_+)(A_\nu^-A^\nu_-)-(A_\mu^+A^\mu_-)^2] \cr
&+&\frac{1}{2}(\partial_\mu\sigma)^2
-\lambda v^2\sigma^2-\lambda v \sigma^3-\frac{\lambda}{4}\sigma^4,
\label{eq:NAH}
\end{eqnarray}
with the charged vector field 
$A_\mu^\pm \equiv \frac{1}{\sqrt{2}}(A_\mu^1 \pm iA_\mu^2) \in {\bf C}$ 
and its mass $M \equiv ev$. 
Using the unbroken U(1) gauge (photon) field $A_\mu\equiv A_\mu^3$,  
we define 
the U(1) covariant derivative $D_\mu^\pm \equiv \partial_\mu \pm ieA_\mu$ and 
the U(1) field strength $F_{\mu\nu}\equiv \partial_\mu A_\nu-\partial_\nu A_\mu$, 
which satisfy $[D_\mu^\pm,D_\nu^\pm]=\pm ieF_{\mu\nu}$. 

In the NAH theory, 
the bilinear part of the charged vector field $A_\mu^\pm$ is expressed by 
\begin{eqnarray}
{\cal L}_{A^\pm {\rm 2nd}}&=&
-\frac{1}{2}(D_\mu^-A_\nu^+-D_\nu^-A_\mu^+)(D^\mu_+ A^\nu_--D^\nu_+ A^\mu_-)
+(M+e\sigma)^2 A_\mu^+A^\mu_- 
+ieF^{\mu\nu}A_\mu^+A_\nu^- \cr
&=&-A_\mu^+D_\nu^+(D^\mu_+ A^\nu_--D^\nu_+ A^\mu_-)
+(M+e\sigma)^2 A_\mu^+A^\mu_- 
+ieF^{\mu\nu}A_\mu^+A_\nu^- \cr
&=&
A_\mu^+[D_\nu^+D^\nu_+ +(M+e\sigma)^2]A^\mu_-
+2ieF^{\mu\nu}A_\mu^+A_\nu^- 
-A_\mu^+D^\mu_+D_\nu^+ A^\nu_-.
\end{eqnarray}
Note here that the term 
``$2ieF^{\mu\nu}A_\mu^+A_\nu^-$'' corresponds to 
the gyromagnetic ratio $g=2$ \cite{NO78,BCGK14}
for the charged vector field $A_\mu^\pm$, which is a general property 
of SU($N$) non-Abelian gauge theories, including the NAH theory and QCD.

\section{Non-Abelian Higgs Theory in a Strong Magnetic Field}

Now, we study the non-Abelian Higgs (NAH) theory 
in the presence of an external field of $A_\mu$ or $F_{\mu\nu}$, e.g., 
a strong magnetic field $B$. 
In this paper, we mainly consider a constant external magnetic field $B(\ge 0)$ 
in the $z$-direction, i.e., $F_{12}=B$.

For the simple treatment, while $eB$ can take a large value,
the gauge coupling $e (>0)$ is taken to be small such that one can drop off 
the $O(e^2)$ self-interaction terms of charged vector fields $A_\mu^\pm$ 
in the NAH Lagrangian (\ref{eq:NAH}). 
Also, $\lambda$ is taken to be enough large such that 
$v$ and $M \equiv ev$ are unchanged under the magnetic field. 

In such a system, only the bilinear term of $A_\mu^\pm$ is important, 
since it includes the coupling with the external photon field $A_\mu$, 
and the $O(e^2)$ self-interaction terms of $A^\pm$ without $A_\mu$ 
and the Higgs fluctuation $\sigma$ can be dropped off 
in the NAH Lagrangian (\ref{eq:NAH}). 
Therefore, the NAH theory is mainly expressed by 
the bilinear term of the charged vector field $A_\mu^\pm$, 
\begin{eqnarray}
{\cal L}_{A^\pm {\rm 2nd}}=A_\mu^+(D_\nu^+D^\nu_+ +M^2)A^\mu_-
+2ieF^{\mu\nu}A_\mu^+A_\nu^- -A_\mu^+D^\mu_+D_\nu^+ A^\nu_-.
\label{eq:CV}
\end{eqnarray}

\subsection{Field equation for the charge vector field}

From the Lagrangian (\ref{eq:CV}), 
the field equations for the charged vector field $A_\mu^\pm$ are given by 
\begin{eqnarray}
(D_\nu^+D^\nu_+ +M^2)A^\mu_-
+2ieF^{\mu\nu}A_\nu^- -D^\mu_+D_\nu^+ A^\nu_-=0, \cr
(D_\nu^-D^\nu_- +M^2)A^\mu_+ 
-2ieF^{\mu\nu}A_\nu^+ -D^\mu_-D_\nu^- A^\nu_+=0.
\label{eq:FE}
\end{eqnarray}
Multiply Eq.(\ref{eq:FE}a) by $D_\mu^+$ from the left  
and using $[D_\mu^\pm,D_\nu^\pm]=\pm ieF_{\mu\nu}$, we obtain 
\begin{eqnarray}
ie[D^\mu_+, F_{\mu\nu}] A^\nu_- + M^2 D_\mu^+ A^\mu_-
=ie (\partial^\mu F_{\mu\nu}) A^\nu_- + M^2 D_\mu^+ A^\mu_-=0.
\end{eqnarray}
For the constant electromagnetic field $F_{\mu\nu}$, 
the massive charged-vector field $A_\mu^\pm$ with a non-zero $M$  
satisfies the maximally Abelian (MA) gauge condition, 
\begin{eqnarray}
D_\mu^+ A^\mu_-=(\partial_\mu+ieA_\mu)A^\mu_-=0,
\label{eq:MA}
\end{eqnarray}
and the field equations (\ref{eq:FE}) become 
\begin{eqnarray}
(D_\nu^\pm D^\nu_\pm +M^2)A^\mu_\mp \pm 2ieF^{\mu\nu}A_\nu^\mp=
[(D_\lambda^\pm D^\lambda_\pm +M^2)g^{\mu\nu} \pm 2ieF^{\mu\nu}]A_\nu^\mp=0.
\end{eqnarray}

For the constant magnetic field $B=F_{12}(\ge 0)$ in the $z$-direction, 
one finds 
\begin{eqnarray}
[(D_\lambda^\pm D^\lambda_\pm +M^2)g_{\mu\nu} 
\pm 2ieB\epsilon_{\mu\nu3}]A^\nu_\mp=
[(D_\lambda^\pm D^\lambda_\pm +M^2)\hat \eta 
+2eB \hat S_z^q]_{\mu\nu}A^\nu_\mp=0,
\end{eqnarray}
with $(\hat \eta)_{\mu\nu}=g_{\mu\nu}$ and 
the spin-charge operator $\hat S_z^q \equiv Q \hat S_z$.
Here, for the charged vector field $A^\mu_\pm$, 
$Q$ is the U(1) charge $Q \in \{\pm1,0\}$ (in the unit of $e$), 
and $(\hat S_z)_{\mu\nu}=-i\epsilon_{\mu\nu3}$
is a spin operator 
and its non-zero eigenvalue states are 
$A^R_\pm \equiv \frac{1}{\sqrt{2}}(A^x_\pm-i A^y_\pm)$ ($s_z=1$) and 
$A^L_\pm \equiv \frac{-i}{\sqrt{2}}(A^x_\pm+iA^y_\pm)$ ($s_z=-1$).
Thus, we obtain  
\begin{eqnarray}
[(D_\lambda^\pm D^\lambda_\pm +M^2)\hat \eta + 2eB \hat S_z^q] {\cal A}_\alpha=0 
\qquad (\alpha = \pm 1, 0)
\label{eq:FEM}
\end{eqnarray}
in the diagonal basis of $\hat S_z^q$ and $\hat \eta$.
${\cal A}_\alpha$ is a linear combination of  $A^\mu_\pm$ 
so as to satisfy 
$\hat S_z^q {\cal A}_\alpha =\alpha {\cal A}_\alpha$: 
\begin{eqnarray}
{\cal A}_{+1} = A^R_+ ~~{\rm or}~~ A^L_- , \qquad
{\cal A}_0 = A^z_\pm ,~A^t_\pm \qquad
{\cal A}_{-1} = A^L_+  ~~{\rm or}~~ A^R_- .
\end{eqnarray}

\subsection{Eigenstates and the Landau level}

Now, we consider the eigenstate of the charged vector field $A_\mu^\pm$ 
in the constant magnetic field, 
by investigating the operator 
in the field equation (\ref{eq:FEM}) for the charged vector field $\vec A_\pm$, 
\begin{eqnarray}
\hat K &\equiv& -(D_\lambda^\pm D^\lambda_\pm +M^2) + 2eB \hat S_z^q
=-\partial_t^2+\partial_z^2
+\{(\partial_x \pm ieA_x)^2+(\partial_y \pm ieA_y)^2\} -M^2 +2eB \hat S_z^q \cr
&=&
\hat p_t^2-\hat p_z^2-2eB\left(\hat a^\dagger \hat a+\frac{1}{2}\right)
-M^2 +2eB \hat S_z^q, 
\end{eqnarray}
where $\hat p_\mu \equiv i \partial_\mu $ and
$\hat a \equiv\frac{1}{\sqrt{2eB}}[(\hat p_x \pm eA_x)-i(\hat p_y \pm eA_y)]$.
Here, $\hat K$ has a one-dimensional harmonic oscillator in the $x$, $y$ direction 
\cite{NO78,ST91}, and we  introduce 
the diagonal basis such as $\hat a^\dagger \hat a|n\rangle =n|n\rangle$. 
Then, the eigenvalue of $\hat K$ is given by
\begin{eqnarray}
K(p_t, p_z, n, s_z) \equiv
p_t^2-p_z^2-2eB\left(n+\frac{1}{2}-s_z\right)-M^2 \quad (n=0,1,2, ... ~;~ s_z=\pm1,0) 
\end{eqnarray}
for the $n$-th Landau level with 
the eigenvalue $s_z$ of $\hat S_z^q$.
The eigenstate $|p_t, p_z, n, s_z \rangle$ which satisfies  
\begin{eqnarray}
\hat K |p_t, p_z, n, s_z \rangle = K(p_t, p_z, n, s_z) |p_t, p_z, n, s_z \rangle
\end{eqnarray}
is expressed in the coordinated space as 
\begin{eqnarray}
\psi_{p_t p_z n s_z}(x) \equiv \langle x^\mu| p_t, p_z, n, s_z \rangle 
=\langle t| p_t \rangle \langle z| p_z \rangle \langle x,y| n \rangle \chi_{s_z}
=e^{-ip_t t}e^{ip_z z}\psi_n(x,y)\chi_{s_z},
\label{eq:ES}
\end{eqnarray}
with the harmonic-oscillator eigenstate $\psi_n(x,y)$ 
and the spin eigenstate $\chi_{s_z}$, satisfying $\hat S_z^q \chi_{s_z}=s_z \chi_{s_z}$.
For the external U(1) gauge (photon) field $A_\mu \equiv A_\mu^3$, 
we take the symmetric gauge, $(A_x, A_y)=\frac{B}{2}(y, -x)$ with $A_t=A_z=0$. 
(Physical results never depend on the remaining U(1) gauge choice.) 
Then, the harmonic-oscillator eigenstate $\psi_n(x,y)$ is written 
in 2-dimensional polar coordinates by
\begin{eqnarray}
\psi_n(x,y)=\frac{1}{l}\sqrt{\frac{n!}{(n-m)!}}
e^{-\frac{r^2}{4l^2}}\left(\frac{r}{\sqrt{2}l}\right)^{|m|}
L_n^{|m|}(\frac{r^2}{2l^2}) \frac{e^{im\varphi}}{\sqrt{2\pi}}  \quad (|m| \le n)
\end{eqnarray}
with $r \equiv \sqrt{x^2+y^2}$ and
the associated Laguerre polynomial $L_n^{m}$.
Here, $l \equiv \frac{1}{\sqrt{eB}}$ is a typical length 
of the minimal cyclotron orbit in the magnetic field $B$.

For the on-mass-shell state of the charged vector field $A_\mu^\pm$ 
satisfying the field equation (\ref{eq:FEM}), 
its energy $p_t$ is given by \cite{NO78} 
\begin{eqnarray}
p_t^2=p_z^2+2eB\left(n+\frac{1}{2}-s_z\right)+M^2 \quad (n=0,1,2, ... ;~ s_z=\pm1,0)
\end{eqnarray}
for the Landau level with $n$ and $s_z$.
In the strong magnetic field, the lowest Landau level (LLL) with $n=0$ and $s_z=1$ 
gives a dominant contribution, and therefore one can use 
the LLL approximation \cite{M96}, 
keeping only the LLL contribution. 
Here, the LLL wave-function is written by
\begin{eqnarray}
\psi_{LLL} (x,y,z) \equiv \psi_0(x,y) e^{ip_zz} \chi_{s_z=1}
=\sqrt{\frac{eB}{2\pi}} e^{-\frac{x^2+y^2}{4l^2}} e^{ip_zz} \chi_{s_z=1},
\end{eqnarray}
which is localized within the order of the length $l \equiv \frac{1}{\sqrt{eB}}$ 
in the $x$, $y$ direction. 
The LLL energy $p_t$ is 
\begin{eqnarray}
p_t^2=p_z^2-eB+M^2=p_z^2-\mu^2, \quad \mu \equiv \sqrt{M^2-eB}.
\end{eqnarray}

\subsection{Propagator of the charge vector field}

From the Lagrangian (\ref{eq:CV}), 
the propagator $\hat D_{\mu\nu}$ of the charged vector field $A_\mu^\pm$ 
is expressed as  
\begin{eqnarray}
\hat D^{-1}_{\mu\nu}=-(D_\lambda^+D^\lambda_+ +M^2)g_{\mu\nu}
-2ieF_{\mu\nu} +D_\mu^+D_\nu^+,
\label{eq:PR}
\end{eqnarray}
which actually satisfies   
${\cal L}_{A^\pm {\rm 2nd}}=-A^\mu_+\hat D^{-1}_{\mu\nu}A^\mu_-.$
In the RHS of Eq.(\ref{eq:PR}), the first two terms are responsible to 
the physical pole of the charged vector field $A^\mu_\pm$, 
and the last term $D_\mu^+D_\nu^+$ 
suffers from gauge transformation. 

Here, we consider addition of a gauge-fixing term whose 
bilinear part of $A^\mu_\pm$ is 
\begin{eqnarray}
{\cal L}_{A^\pm}^{\rm g.f.}=-\frac{1}{\alpha}
(D_\mu^-A^\mu_+)
(D_\nu^+A^\nu_-)=
+\frac{1}{\alpha}A_\mu^+D^\mu_+D_\nu^+ A^\nu_-.
\end{eqnarray}
By taking $\alpha=1$, 
the total bilinear part of the charged vector field $A^\mu_\pm$ is expressed as 
\begin{eqnarray}
{\cal L}_{A^\pm {\rm 2nd}}^{\rm tot} 
\equiv {\cal L}_{A^\pm {\rm 2nd}}+{\cal L}_{A^\pm}^{\rm g.f.}
=A_\mu^+(D_\nu^+D^\nu_+ +M^2)A^\mu_-+2ieF^{\mu\nu}A_\mu^+A_\nu^-,
\end{eqnarray}
and the charged-vector propagator $\hat D_{\mu\nu}$ becomes 
\begin{eqnarray}
\hat D^{-1}_{\mu\nu}=-(D_\lambda^+D^\lambda_+ +M^2)g_{\mu\nu}
-2ieF_{\mu\nu}.
\end{eqnarray}

For the constant magnetic field $B=F_{12}$, $\hat D_{\mu\nu}$ is written 
with $\hat\eta_{\mu\nu}=g_{\mu\nu}$ and
$(\hat S_z^q)_{\mu\nu}=i\epsilon_{\mu\nu3}$ as
\begin{eqnarray}
\hat D^{-1}_{\mu\nu}
=-(D_\lambda^+D^\lambda_+ +M^2)g_{\mu\nu}
-2eBi \epsilon_{\mu\nu3} 
=-[(D_\lambda^+D^\lambda_+ +M^2)\hat \eta +2eB \hat S_z^q]_{\mu\nu}.
\end{eqnarray}
The charged-vector propagator $\hat D$ has 
the space-time structure of 
\begin{eqnarray}
\hat D^{-1} &\equiv& -[(D_\lambda^+D^\lambda_+ +M^2)\hat \eta 
+2eB \hat S_z^q] 
=[\hat p_t^2-\hat p_z^2
-2eB( \hat a^\dagger \hat a+1/2 )-M^2] \hat \eta
-2eB\hat S_z^q \cr
&=&
\left(
    \begin{array}{cccc}
      -\hat K_0 &  & & \\
       &  \hat K_0 & -2eBi & \\
         & 2eBi& \hat K_0 &   \\
       & & & \hat K_0
    \end{array}
  \right) 
=
\Omega \left(
    \begin{array}{cccc}
      -\hat K_0 &  & & \\
       &  \hat K_0+2eB & & \\
         & & \hat K_0-2eB &   \\
       & & & \hat K_0
    \end{array}
  \right) \Omega^\dagger
\equiv\Omega \hat {\cal D}^{-1} \Omega^\dagger~~~~~~~
\end{eqnarray}
with $\hat K_0 \equiv D_\lambda^+D^\lambda_+ +M^2$ and 
the spatial rotation  
$
  \Omega \equiv \left(
    \begin{array}{cccc}
1 & & & \\
&    \frac1{\sqrt{2}} & \frac{-i}{\sqrt{2}}  & \\
&    \frac{-i}{\sqrt{2}}& \frac{1}{\sqrt{2}} & \\
& & & 1
    \end{array}
  \right)
$
which diagonalizes $\hat D$.
By the spatial rotation $\Omega$, the coordinate basis  
$(\hat t, \hat x, \hat y, \hat z)$ is transformed to 
the $\hat D$-diagonalized basis 
$(\hat t, \hat R, \hat L, \hat z)^T=\Omega(\hat t, \hat x, \hat y, \hat z)^T$. 
Note that only one spatial sector of 
$\hat L \equiv\frac{-i}{\sqrt{2}}(\hat x+i \hat y)$-component in $\hat {\cal D}$
includes the lowest Landau level.
For the spatial sector $\hat {\cal D}_s$ in 
the diagonalized charged-vector propagator $\hat {\cal D}$, 
its eigenvalue $D_s(p_t, p_z, n, s_z)$ is written by
\begin{eqnarray}
D_s^{-1}(p_t, p_z, n, s_z) \equiv 
\langle p_t, p_z, n, s_z|\hat {\cal D}_s^{-1}| p_t, p_z, n, s_z \rangle
=-p_t^2+p_z^2+2eB \left( n+\frac12-s_z \right)+M^2.~~~~~~
\end{eqnarray}
Then, the coordinate-space representation of 
the spatial charged-vector propagator $\hat {\cal D}_s$ is given by
\begin{eqnarray}
&&D_s(x, x')  \equiv \langle x|\hat {\cal D}_s|x' \rangle=
\sum_{p_t, p_z, n, s_z} \langle x| p_t, p_z, n, s_z \rangle 
D_s(p_t, p_z, n, s_z)  \langle p_t, p_z, n, s_z |x' \rangle \cr
&=& \int_{-\infty}^{\infty} \frac{dp_t}{2\pi} \int_{-\infty}^{\infty} \frac{dp_z}{2\pi} 
\sum_{n=0}^\infty \sum_{s_z=-1}^1 
\psi_{p_t p_z n s_z}(x) \frac{1}{-p_t^2+p_z^2+2eB(n+\frac12-s_z)
+M^2}\psi_{p_t p_z n s_z}^\dagger(x')~~~~~~~~~~
\end{eqnarray}
with the eigenstate 
$\psi_{p_t p_z n s_z}(x) \equiv \langle x^\mu| p_t, p_z, n, s_z \rangle$ 
in Eq.(\ref{eq:ES}).

In the strong constant magnetic field, we use the 
lowest Landau level (LLL) approximation, 
keeping only the LLL ($n=0$, $s_z=1$) contribution: 
\begin{eqnarray}
D_s(x, x') &\simeq& D_{LLL}(x, x') \equiv \sum_{p_t, p_z, n, s_z}
\langle x|p_t, p_z, n, s_z \rangle D_s(p_t, p_z, n, s_z) \langle p_t, p_z, n, s_z|x' \rangle 
\delta_{n 0} \delta_{s_z 1} \cr
&=&\int_{-\infty}^\infty \frac{dp_t}{2\pi} \int_{-\infty}^\infty \frac{dp_z}{2\pi} 
e^{-ip_t(t-t')}e^{ip_z(z-z')}
\psi_0(x,y) \psi_0^\dagger(x',y') \frac{1}{-p_t^2+p_z^2-eB+M^2} \cr
&=&\frac{eB}{2\pi}~e^{-\frac{x^2+y^2}{4l^2}} e^{-\frac{x'^2+y'^2}{4l^2}}
\int_{-\infty}^\infty \frac{dp_t}{2\pi} \int_{-\infty}^\infty \frac{dp_z}{2\pi} 
e^{-ip_t(t-t')}e^{ip_z(z-z')}\frac{-1~~}{p_t^2-p_z^2-\mu^2},
\label{eq:LLL}
\end{eqnarray}
with $\mu^2 \equiv M^2-eB$ and $l \equiv \frac{1}{\sqrt{eB}}$.
The factor $\frac{eB}{2\pi}$ corresponds to the Landau degeneracy.

This spatial propagator $D_s(x,x')\simeq D_{LLL}(x,x')$ 
of the charged vector field $A_\mu^\pm$ 
is similar to a 1+1 dimensional propagator with the mass $\mu$, 
although $D_{LLL}(x,x')$ acts on the spatial ($\hat L$) component of the charge current
and hence the total sign is different.
In fact, in the strong constant magnetic field, 
the charge motion in the $x$, $y$ direction is frozen, and  
the low-dimensionalization is realized.
Actually, the correlation brought by $D_{LLL}(x,x')$ in Eq.(\ref{eq:LLL}) is localized 
within the order of the length $l \equiv \frac{1}{\sqrt{eB}}$ in the $x$, $y$ direction. 

When the strong constant magnetic field $eB$ is 
near and below the critical value $eB_c \equiv M^2$, 
the effective mass-squared $\mu^2 \equiv M^2-eB$ 
becomes small (and non-negative), 
and the charged vector fields $A_\mu^\pm$ 
behave as 1+1-dimensional quasi-massless fields, 
and therefore 
this spatial propagator $D_{LLL}(x,x')$ 
induces a strong correlation along the magnetic-field direction 
between off-diagonal charges coupled with $A_\mu^\pm$. 
(For $\mu^2<0$, i.e., $eB>M^2$, the Nielsen-Olesen instability occurs 
\cite{NO78, BCGK14}.)

For the off-diagonal current $J^\mu_\pm$ coupled with $A^\mu_\mp$, 
the current-current correlation is derived as 
\begin{eqnarray}
S_{JJ}=\int d^4x~d^4x'~ J^\mu_+(x) \hat D_{\mu\nu}(x,x') J^\nu_-(x')
\simeq \int d^4x~d^4x'~ J^{LLL}_+(x) \hat D_{LLL}(x,x') J^{LLL}_-(x'),
\end{eqnarray}
where $J^{LLL}_\pm$ is the spatial component coupled 
with the LLL state of $A^\mu_\mp$.  
The inter-charge potential $V(r)$ along $\vec B$ is estimated 
with a spiral current 
$J^{LLL}_\pm \sim Q_{\pm}\delta(x)\delta(y)\delta(z\pm\frac{r}{2})$ localized near $x=y=0$,
\begin{eqnarray}
V(r)=-Q_+Q_- \frac{eB}{4\pi}\frac{1}{\mu}e^{-\mu r} 
\quad \rightarrow \quad V(r)=Q_+Q_-\frac{eB}{4\pi}r 
\quad {\rm as} \quad eB \rightarrow eB_c \equiv M^2.
\end{eqnarray}
In the limit of 
$eB \rightarrow eB_c -0$, the charged vector fields $A^\mu_\pm$ become massless 
as $\mu^2 \equiv M^2-eB \rightarrow +0$, and 
the linear potential appears along $\vec B$. 
(This strong correlation may relate to the Nilsen-Olesen instability.)
In this system, the Landau degeneracy $\frac{eB}{2\pi}$ gives 
the dimension of the string tension.

Then, for example, for the fermion (fundamental-rep.) coupled with the SU(2) gauge field, 
the charged fermion makes the cyclotron motion 
in the strong constant magnetic field $\vec B$,
with suffering from 
the strong correlation along $\vec B$ induced 
by charged vector fields $A_\mu^\pm$ (see Fig.1).

\begin{figure}[h]
\centering
\includegraphics[height=4cm]{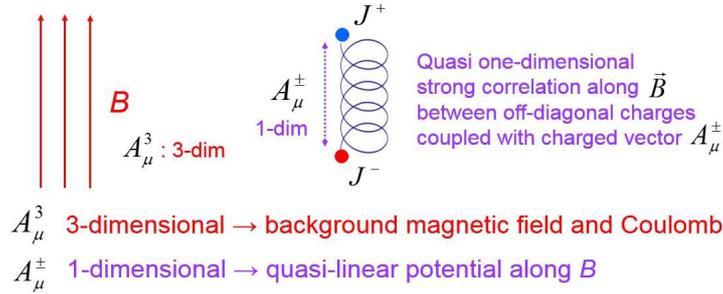}
\caption{
The SU(2) NAH theory in the strong constant magnetic field $\vec B$ 
near and below $eB_c \equiv M^2$. 
The 3-dimensional massless photon $A_\mu \equiv A_\mu ^3$ 
gives the background magnetic field $\vec B$, and induces 
the cyclotron motion for charged particles. 
The charged vector fields $A_\mu^\pm$ become spatially quasi-one-dimensional, 
and induce a strong correlation along $\vec B$ 
between off-diagonal charges coupled with $A_\mu^\pm$.
}
\end{figure}

\section{Summary and Conclusion}

We have investigated the non-abelian Higgs (NAH) theory
with the Higgs triplet in a strong constant magnetic field $\vec B$, 
where the lowest-Landau-level (LLL) approximation can be used. 
We have found that, near and below the critical magnetic field of $eB_c \equiv M^2$, 
the charged vector fields $A_\mu^\pm$ 
behave as spatially one-dimensional quasi-massless fields, 
and give a strong correlation along $\vec B$ 
and eventually a linear potential $V(r) \propto eBr$ at $eB=eB_c$ 
between off-diagonal charges coupled with $A_\mu^\pm$. 
This may lead a new type of confinement 
caused by charged vector fields $A_\mu^\pm$. 

For this confinement, although the external photon field $A_\mu$ 
is important for the formation the LLL state of $A_\mu^\pm$,  
off-diagonal charged-vector fields $A_\mu^\pm$ 
plays an essential role to the linear confinement potential $V(r) \propto eBr$, 
so that this can be called ``off-diagonal (charged) dominance'' (see Fig.2), 
in contrast to ``Abelian dominance'' for quark confinement in QCD in the MA gauge.

\begin{figure}[h]
\centering
\includegraphics[height=5cm]{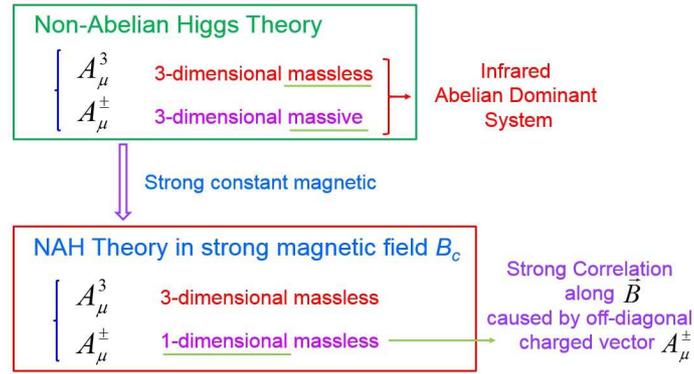}
\caption{
The SU(2) NAH theory with and without the magnetic field.
At $B=0$, the Higgs mechanism gives a large mass $M$ for $A_\mu^\pm$ 
and infrared Abelian dominance.
At the critical magnetic field of $eB_c = M^2$, however, 
off-diagonal charged-vector fields $A_\mu^\pm$ 
become spatially one-dimensional and massless, 
and play an essential role to the linear potential 
$V(r) \propto eBr$
between off-diagonal charges coupled with $A_\mu^\pm$. 
}
\end{figure}

\begin{acknowledgments}
I thank Prof. K. Konishi for his useful discussions on 
the relation to non-Abelian vortices near the Nielsen-Olesen instability.
I also thank Prof. M.N. Chernodub for his useful comments on the case of 
the Weinberg-Salam model.
This work is supported in part by the Grants-in-Aid  for Scientific Research 
[15K05076] from Japan Society for the Promotion of Science.
\end{acknowledgments}


\begin{thebibliography}{50}

\bibitem{EBH64}
F.~Englert and R.~Brout, {\it Phys. Rev. Lett.} {\bf 13}, 321 (1964); \\
P.W.~Higgs, {\it Phys. Lett.} {\bf 12}, 132 (1964), 
{\it Phys. Rev. Lett.} {\bf 13}, 508 (1964). 

\bibitem{N74}
Y.~Nambu, {\it Phys. Rev.} {\bf D10}, 4262 (1974);
G.~'t~Hooft, in {\it High Energy Physics}, 
(Editorice Compositori, Bologna, 1975); 
S.~Mandelstam, {\it Phys. Rept.} {\bf 23}, 245 (1976).

\bibitem{tH81} G.~'t~Hooft, 
{\it Nucl. Phys.} {\bf B190}, 455 (1981).

\bibitem{EI82} Z.~F.~Ezawa and A.~Iwazaki, 
{\it Phys. Rev.} {\bf D25}, 2681 (1982).

\bibitem{AS99} K.~Amemiya and H.~Suganuma, 
{\it Phys. Rev.} {\bf D60}, 114509 (1999); 
S. Gongyo and H. Suganuma, 
{\it Phys. Rev.} {\bf D87}, 074506 (2013);
S.~Gongyo, T.~Iritani and H.~Suganuma, 
{\it Phys. Rev.} {\bf D86}, 094018 (2012).

\bibitem{SY90} T.~Suzuki and I.~Yotsuyanagi, 
{\it Phys. Rev.} {\bf D42}, 4257(R) (1990); \\
J.D.~Stack, S.D.~Neiman, and R.J.~Wensley, 
{\it Phys. Rev.} {\bf D50}, 3399 (1994); \\
O.~Miyamura, {\it Phys. Lett.} {\bf B353}, 91 (1995); 
R.M.~Woloshyn, {\it Phys. Rev.} {\bf  D51}, 6411 (1995).

\bibitem{SS1415} N.~Sakumichi and H.~Suganuma, 
{\it Phys. Rev.} {\bf D90}, 111501(R) (2014); 
{\it Phys. Rev.} {\bf D92}, 034511 (2015).

\bibitem{S77}
G.K.~Savvidy, {\it Phys. Lett.} {\bf 71B}, 133 (1977).

\bibitem{NO78}
N.K.~Nielsen and P.~Olesen, {\it Nucl. Phys.} {\bf B144}, 376 (1978). 

\bibitem{AO89}
J.~Ambjorn and P.~Olesen, {\it Nucl. Phys.} {\bf B170}, 265 (1980); 
{\it Nucl. Phys.} {\bf B315}, 606 (1989).

\bibitem{ST91}
H.~Suganuma and T.~Tatsumi, {\it Ann. Phys.} {\bf 208}, 470 (1991);
{\it Phys. Lett.}  {\bf B269}, 371  (1991); 
{\it Prog. Theor. Phys.} {\bf 90}, 379 (1993). 

\bibitem{M96}
V.P.~Gusynin, V.A.~Miransky and I.A.~Shovkovy, 
{\it Nucl. Phys.} {\bf  B462}, 249 (1996); \\
For a recent review, V.A.~Miransky and I.A.~Shovkovy, 
{\it Phys. Rept.} {\bf 576}, 1 (2015).

\bibitem{KMW08}
D.E.~Kharzeev, L.D.~McLerran and H.J.~Warringa, 
{\it Nucl. Phys.} {\bf A803}, 227 (2008).

\bibitem{CDV13}
M.N.~Chernodub, J.V.~Doorsselaere and H.~Verschelde, 
{\it Phys. Rev.} {\bf D88}, 065006 (2013). 

\bibitem{BCGK14}
S.~Bolognesi, C.~Chatterjee, S.B.~Gudnason and K.~Konishi, 
{\it JHEP} {\bf 1410}, 101 (2014).

\end{thebibliography}
\end{document}